\documentclass[aps,pra,preprint,superscriptaddress,showpacs,endfloats*]{revtex4}
\usepackage{graphicx}
\usepackage{dcolumn}
\usepackage{bm}
\usepackage{amsfonts}
\usepackage{amsmath}

\begin{document}
\baselineskip=0.5cm
\renewcommand{\thefigure}{\arabic{figure}}

\title{Density-functional theory of strongly correlated Fermi 
gases in elongated harmonic traps}
\author{Gao Xianlong}
\affiliation{NEST-CNR-INFM and Scuola Normale Superiore, I-56126 Pisa, Italy}
\author{Marco Polini}
\email{m.polini@sns.it}
\affiliation{NEST-CNR-INFM and Scuola Normale Superiore, I-56126 Pisa, Italy}
\author{Reza Asgari}
\affiliation{Institute for Studies in Theoretical Physics and Mathematics, Tehran 19395-5531, Iran and NEST-CNR-INFM}
\author{M. P. Tosi}
\affiliation{NEST-CNR-INFM and Scuola Normale Superiore, I-56126 Pisa, Italy}
\date{\today}

\begin{abstract}
Two-component Fermi gases with tunable repulsive or attractive interactions 
inside quasi-one-dimensional (Q$1D$) harmonic wells may soon become the cleanest laboratory realizations 
of strongly correlated Luttiger and Luther-Emery liquids under confinement.
We present a microscopic Kohn-Sham density-functional theory of these systems, with specific attention to a gas on the approach to a confinement-induced Feshbach resonance. The theory employs the one-dimensional 
Gaudin-Yang model as the reference system and transfers the appropriate Q$1D$ 
ground-state correlations to the confined inhomogeneous gas {\it via} a suitable 
local-density approximation to the exchange and correlation energy functional. 
Quantitative understanding of the role of the interactions 
in the bulk shell structure of the axial density profile is thereby achieved. 
While repulsive intercomponent interactions depress the amplitude of the shell structure of the noninteracting gas, 
attractive interactions stabilize atomic-density waves through spin pairing. These 
should be clearly observable in atomic clouds containing of the order of up to a hundred atoms.
\end{abstract}
\maketitle

\section{Introduction}
\label{sect:intro}

Ultracold atomic gases, which are highly tunable and ideally clean, are 
attracting a great deal of interdisciplinary interest. In particular their study may help us 
understand a number of phenomena that have been predicted in solid-state and condensed-matter physics~\cite{cirac_2003}. 
Several effects, known in these subfields of physics for decades, 
have already been observed and quantitatively analyzed in ultracold atomic gases. 
Three beautiful examples are the Bloch oscillations under an applied force 
in a one-dimensional ($1D$) optical lattice~\cite{bloch_oscillations}, 
the formation of highly ordered Abrikosov lattices of vortices in rapidly rotating harmonic traps~\cite{rotatingcoldatoms}, and the superfluid-to-Mott insulator transition of a condensate in a $3D$ 
optical lattice~\cite{smi}.

Cold atoms have also been successfully trapped 
in low-dimensional geometries~\cite{cold_atoms_low_D}. 
Typical $1D$ quantum phenomena 
have already been observed in both Bose and Fermi gases. 
For instance, in the work of Paredes {\it et al.}~\cite{cold_atoms_low_D} and of Kinoshita {\it et al.}~\cite{cold_atoms_low_D} a $^{87}{\rm Rb}$ gas has been used to realize experimentally 
a Tonks-Girardeau system~\cite{TG_gas}. The more recent preparation of two-component Fermi gases 
in a quasi-$1D$ (Q$1D$) geometry~\cite{moritz_2005} provides a unique possibility to experimentally study phenomena that were predicted a long time ago for electrons in a $1D$ solid-state environment, 
such as spin-charge separation in Luttinger liquids~\cite{luttinger_liquids,Giuliani_and_Vignale} and charge-density waves in Luther-Emery liquids~\cite{luther_emery}. The experiment by Moritz {\it et al.}~\cite{moritz_2005} also 
offers the opportunity of testing a Q$1D$ integrable model of the BCS-BEC crossover~\cite{fuchs_2004,tokatly_2004}, which is based on the idea of a confinement-induced resonance (CIR)~\cite{olshanii_1998}. 

In~\cite{fuchs_2004,tokatly_2004} the gas has been assumed to be translationally invariant along the axial direction, and thus the authors have been able to provide an analytical description of the crossover by employing the exact Bethe-Ansatz (BA) solution of the homogenous Gaudin-Yang model (HGYM)~\cite{GY} and of the homogeneous Lieb-Liniger model~\cite{LL}. The present work focuses instead on inhomogeneous Q$1D$ Fermi gases inside highly elongated harmonic traps and treats the axial confinement by means of the Hohenberg-Kohn-Sham density-functional theory (DFT)~\cite{dft,Giuliani_and_Vignale}. With a few exceptions~\cite{general,burke_2004,lima_2002,gao_2005}, 
most applications of DFT have used so far as the underlying reference fluid the homogeneous 
electron gas, which is a normal Fermi liquid over a wide range of density. In our present study we use the HGYM as the reference fluid, in order to transfer to the inhomogeneous gas the Luttinger and Luther-Emery $1D$ correlations. 

It is appropriate at this point to refer to related theoretical studies dealing with Q$1D$ inhomogeneous Fermi gases~\cite{gao_2002,recati_2003,astrakharchik_2004,kecke_2005}. In Ref.~\cite{gao_2002} a bosonization technique has been used to calculate analytically the density profile, the momentum distribution, and several correlation functions of two-component Fermi gases with inclusion of intercomponent forward-scattering processes. In Refs.~\cite{recati_2003,astrakharchik_2004,kecke_2005} the Thomas-Fermi approximation (see the discussion in Sect.~\ref{sect:ks_dft} below) and the so-called inhomogeneous Tomonaga-Luttinger liquid model 
have been used to calculate the density profile of a large system and to discuss spin-charge separation in two-component Fermi gases. In the present work we perform microscopic calculations of the ground-state (GS) density profile of systems with arbitrary size, without having to assume neither peculiar intercomponent interactions 
(as in the bosonization scheme of Ref.~\cite{gao_2002}) nor very large atom numbers (as in Refs.~\cite{recati_2003,astrakharchik_2004,kecke_2005}). We give a fully quantitative study of how 
exchange and correlations modify the bulk shell structure of the axial density profile. 
In particular we show that for sufficiently strong attractive interactions, experimentally detectable atomic-density waves (ADWs) are formed by spin pairing along the axial direction, which should be clearly observable in systems with a relatively low number of atoms ($N_{\rm f}\lesssim 100$). These the oversimplified Thomas-Fermi treatments cannot predict.

The contents of the paper are briefly as follows. In Sect.~\ref{sect:theory} we introduce the 
Hamiltonian that we have used to describe the system of present physical interest, 
summarize the properties of the model in the absence of external potentials, and 
describe the self-consistent DFT scheme that we have used to deal with the inhomogeneity. 
In Sect.~\ref{sect:numerical_results} we report and discuss our main numerical results and finally 
in Sect.~\ref{sect:conclusions} we draw our main conclusions. An Appendix 
contains the exact solution of the inhomogeneous model for two atoms only, which is used in the main text for a test of the local density approximation in the extreme limit of low particle numbers.

\section{Theoretical Approach}
\label{sect:theory}

We consider a two-component Fermi gas with $N_{\rm f}$ atoms confined inside a strongly elongated harmonic trap.
The two species of fermionic atoms are assumed to have the same mass $m$ and different pseudospin $\sigma$, $\sigma=\uparrow$ or $\downarrow$. The trapping potential is axially symmetric and is characterized by angular frequencies $\omega_\perp$ and $\omega_\|$ in the radial and the longitudinal directions, with $\omega_\| \ll \omega_\perp$. Correspondingly we introduce the harmonic-oscillator lengths $a_\perp=\sqrt{\hbar/(m\omega_\perp)}$ and $a_{\|}=\sqrt{\hbar/(m\omega_{\|})}$.

The gas is dynamically $1D$ if the anisotropy parameter of the trap is much smaller than the inverse atom number, $\omega_{\|}/\omega_\perp\ll N^{-1}_{\rm f}$. It can thus be described by the Hamiltonian~\cite{astrakharchik_2004}
\begin{equation}\label{eq:igy}
{\cal H}=-\frac{\hbar^2}{2m}\sum_{i=1}^{N_{\rm f}}\frac{\partial^2}{\partial z^2_i}+g_{\rm \scriptscriptstyle 1D}\sum_{i=1}^{N_{\uparrow}}\sum_{j=1}^{N_{\downarrow}}\delta(z_i-z_j)+V_{\rm ext}\,,
\end{equation}
neglecting intracomponent $p$-wave interactions. Here,
\begin{equation}\label{eq:cir}
g_{\rm \scriptscriptstyle 1D}=\frac{2\hbar^2 a_{\rm \scriptscriptstyle 3D}(B)}{ma^2_\perp}\frac{1}{1-{\cal A}a_{\rm \scriptscriptstyle 3D}(B)/a_\perp}
\end{equation}
is the effective $1D$ Olshanii coupling parameter~\cite{olshanii_1998}, with ${\cal A}=|\zeta(1/2)|/\sqrt{2}\simeq 1.0326$ and $\zeta(x)$ being the Hurwitz zeta function, and $V_{\rm ext}=\sum_{i=1}^{N_{\rm f}}
V_{\rm ext}(z_i)=(m\omega^2_{\|}/2)\sum_{i=1}^{N_{\rm f}}z^2_i$ is the external static potential associated 
with the axial confinement. The $3D$ scattering length $a_{\rm \scriptscriptstyle 3D}$ can be tuned by means of a magnetic field $B$ and has the resonant structure $a_{\rm \scriptscriptstyle 3D}(B)=a_{\rm bg}[1-\delta B/(B-B_{\rm F})]$, $B_{\rm F}$ being the position of a Feshbach resonance, $\delta B$ its width, and $a_{\rm bg}$ the so-called background scattering length~\cite{moritz_2005}. 

Choosing the harmonic-oscillator length $a_{\|}$ as unit of length and 
the harmonic-oscillator quantum $\hbar\omega_{\|}$ as unit of energy, 
the Hamiltonian ({\ref{eq:igy}) can be shown to be governed by 
the dimensionless coupling parameter 
\begin{equation}\label{eq_lambda}
\lambda=\frac{g_{\rm \scriptscriptstyle 1D}}{a_{\|}\hbar\omega_{\|}}\,.
\end{equation} 
The parameter $\lambda$ diverges at the CIR, {\it i.e.} 
when the external magnetic field takes the value $B^\star=B_{\rm F}-\delta B
(a^\star_{\rm \scriptscriptstyle 3D}/a_{\rm bg}-1)^{-1}$ with $a^\star_{\rm \scriptscriptstyle 3D}=a_\perp/{\cal A}$. The coupling parameter is negative for $B^\star<B<B_{\rm F}+\delta B$ and positive everywhere else. At the $3D$ Feshbach resonance, {\it i.e.} when $B=B_{\rm F}$, the coupling parameter has the finite value 
$\lambda^{\rm F}=-2 a_{\|}/({\cal A}a_\perp)$. In Fig.~\ref{fig:one} we show the dependence of $\lambda$ on the magnetic field $B$. 

For $B>B^\star$ and $V_{\rm ext}=0$ the Hamiltonian (\ref{eq:igy}) reduces to the HGYM, which can be solved exactly by means of the BA technique for both repulsive ($g_{\rm \scriptscriptstyle 1D}>0$) and attractive ($g_{\rm \scriptscriptstyle 1D}<0$) interactions~\cite{GY}. In the thermodynamic limit ($N_{\rm f},L\rightarrow \infty$, $L$ being the system size) and for a pseudospin-compensated system ($N_\uparrow=N_\downarrow$), the properties of the HGYM are determined by the linear density $n=N_{\rm f}/L$ and by the effective coupling $g_{\rm \scriptscriptstyle 1D}$. These can be conveniently combined into a single dimensionless parameter $\gamma=mg_{\rm \scriptscriptstyle 1D}/(\hbar^2 n)$. 

The energy per atom can be written in terms of the ``momentum distribution" $\rho(k)$ as
\begin{equation}\label{eq:energy_atom}
\varepsilon_{\rm \scriptscriptstyle GS}(n,g_{\rm \scriptscriptstyle 1D})=\frac{\varepsilon_{\rm b}}{2}+\frac{2\pi\nu}{n}\int_{-Q}^{+Q}\frac{dk}{2\pi}\,\frac{\hbar^2 k^2}{2m}\rho(k)\,,
\end{equation}
where $\varepsilon_{\rm b}=0,\nu=1$ for $g_{\rm \scriptscriptstyle 1D}>0$ and $\varepsilon_{\rm b}=-mg^2_{\rm \scriptscriptstyle 1D}/(4\hbar^2),\nu=2$ for $g_{\rm \scriptscriptstyle 1D}<0$. The function $\rho(k)$ can be calculated by solving the Gaudin-Yang BA integral equation~\cite{GY},
\begin{equation}\label{eq:gaudin_yang}
\rho(k)=\frac{\nu}{2\pi}+\frac{\nu}{\gamma n}\int_{-Q}^{+Q}\frac{dq}{2\pi}\,
{\cal K}\left(2(k-q)/(\gamma n)\right)\rho(q)\,,
\end{equation}
where $Q$ is determined by the normalization condition
\begin{equation}\label{eq:normalization}
\int_{-Q}^{+Q}\rho(k)dk=\frac{n}{\nu}
\end{equation}
and the kernel ${\cal K}(x)$ is given by
\begin{equation}
{\cal K}(x)=
\left\{
\begin{array}{ll}
{\displaystyle \int_{-\infty}^{+\infty}dy\,\frac{{\rm sech}{(\pi y/2)}}{[1+(x+y)^2]}} &{\rm for}\,g_{\rm \scriptscriptstyle 1D}>0\\
{\displaystyle \frac{1}{1+x^2/4}}&{\rm for}\,g_{\rm \scriptscriptstyle 1D}<0
\end{array}
\right.
\,.
\end{equation}
For $g_{\rm \scriptscriptstyle 1D}>0$ the HGYM describes a Luttinger liquid~\cite{luttinger_liquids,Giuliani_and_Vignale}, while for $g_{\rm \scriptscriptstyle 1D}<0$ 
it describes a Luther-Emery liquid~\cite{luther_emery}. 

Before proceeding to discuss the properties of the inhomogeneous model under confinement, we should 
stress that Eq.~(\ref{eq:energy_atom})-(\ref{eq:normalization}) describe the homogeneous limit of the model only for $B>B^\star$, {\it i.e.} before the CIR. After the CIR, as discussed in Refs.~\cite{fuchs_2004,tokatly_2004}, the fermion pairs become unbreakable spin-singlet dimers, behaving like bosons with mass $2m$ and density $n/2$. Thus the appropriate homogeneous limit for $B<B^\star$ is the Lieb-Liniger gas of interacting bosons~\cite{LL}, 
and one should resort in treating inhomogeneity to a DFT approach 
such as that proposed by Griffin~\cite{general} (see also the work of Oliveira 
{\it et al.}~\cite{general}).

\subsection{Density-functional theory of Q$1D$ gas in the Kohn-Sham scheme}
\label{sect:ks_dft}

In the presence of a longitudinal external potential the Hamiltonian ${\cal H}$ in Eq.~(\ref{eq:igy}) cannot be diagonalized exactly. We calculate the GS properties of ${\cal H}$ for $B>B^\star$ by resorting to the 
fermionic DFT scheme~\cite{dft,Giuliani_and_Vignale}. 

Within the Kohn-Sham version of 
DFT the GS density, $n_{\rm \scriptscriptstyle GS}(z)=\langle{\rm GS}|\sum_i\delta(z-z_i)|{\rm GS}\rangle$, can be calculated by solving self-consistently the Kohn-Sham-Schr\"odinger (KSS) equations~\cite{dft},
\begin{equation}\label{eq:kss}
\left[-\frac{\hbar^2}{2m}\frac{\partial^2}{\partial z^2}+V_{\rm KS}(z;[n_{\rm \scriptscriptstyle 
GS}(z)])\right]\varphi_\alpha(z)=\varepsilon_\alpha\varphi_\alpha(z)
\end{equation}
with $V_{\rm KS}(z;[n_{\rm \scriptscriptstyle GS}(z)])=v_{\rm H}(z;[n_{\rm \scriptscriptstyle GS}(z)])+v_{\rm xc}(z;[n_{\rm \scriptscriptstyle GS}(z)])+V_{\rm ext}(z)$, together with the closure
\begin{equation}\label{eq:closure}
n_{\rm \scriptscriptstyle GS}(z)=\sum_{\alpha, {\rm occ.}}\Gamma_\alpha\left|\varphi_\alpha(z)\right|^2\,.
\end{equation}
Here the sum runs over the occupied orbitals and the degeneracy factors $\Gamma_\alpha$ satisfy the sum rule 
$\sum_\alpha \Gamma_\alpha=N_{\rm f}$. The first term in the effective Kohn-Sham potential $V_{\rm KS}$ is the Hartree term $v_{\rm H}=g_{\rm \scriptscriptstyle 1D}n_{\rm \scriptscriptstyle GS}(z)$, while the second term is the exchange-correlation (xc) potential defined as the functional derivative of the xc energy ${\cal E}_{\rm xc}[n(z)]$ evaluated at the GS density profile, $v_{\rm xc}=\delta {\cal E}_{\rm xc}[n(z)]/\delta n(z)|_{\rm \scriptscriptstyle GS}$. The total GS energy of the system is given by 
\begin{eqnarray}
{\cal E}_{\rm \scriptscriptstyle GS}[n_{\rm \scriptscriptstyle GS}(z)]&=&\sum_\alpha\Gamma_\alpha\varepsilon_\alpha-\int_{-\infty}^{+\infty}dz\,
v_{\rm xc}(z;[n_{\rm \scriptscriptstyle GS}(z)])n_{\rm \scriptscriptstyle GS}(z)\nonumber\\
&-&\frac{g_{\rm \scriptscriptstyle 1D}}{2}\int_{-\infty}^{+\infty}n^2_{\rm \scriptscriptstyle GS}(z)dz
+{\cal E}_{\rm xc}[n_{\rm \scriptscriptstyle GS}(z)]\,.
\end{eqnarray}

Equations~(\ref{eq:kss}) and~(\ref{eq:closure}) provide a formally exact scheme to calculate $n_{\rm \scriptscriptstyle GS}(z)$ and ${\cal E}_{\rm \scriptscriptstyle GS}$, but ${\cal E}_{\rm xc}$ and $v_{\rm xc}$ need to be approximated. The local-density approximation (LDA) has been shown to provide an excellent description of the GS properties of a variety of inhomogeneous systems~\cite{dft,Giuliani_and_Vignale}. In the following we employ a BA-based LDA (BALDA) functional~\cite{lima_2002,gao_2005} for the xc potential,
\begin{equation}\label{eq:balda}
v^{\rm \scriptscriptstyle BALDA}_{\rm xc}(z;[n_{\rm \scriptscriptstyle GS}(z)])=\left.v^{\rm hom}_{\rm xc}(n,g_{\rm \scriptscriptstyle 1D})\right|_{n\rightarrow n_{\rm \scriptscriptstyle GS}(z)}\,.
\end{equation}
Here the xc potential of the HGYM is defined by 
\begin{equation}\label{eq:vxc_hom}
v^{\rm hom}_{\rm xc}(n,g_{\rm \scriptscriptstyle 1D})
=\frac{\partial}{\partial n} 
\left[n\varepsilon_{\rm \scriptscriptstyle GS}(n,g_{\rm \scriptscriptstyle 1D})
-n\kappa(n)\right]-ng_{\rm \scriptscriptstyle 1D}\,,
\end{equation}
$\kappa(n)=\pi^2\hbar^2n^2/(24m)$ being 
the kinetic energy of the noninteracting gas per atom.

Before discussing specific calculations of the xc potential of the HGYM, several important remarks are in order at this point:
 
(i) In the limit $\lambda=0$ the KSS equations correctly yield the GS density profile of a noninteracting paramagnetic Fermi gas,
\begin{equation}\label{eq:non_interacting}
\left.n_{\rm \scriptscriptstyle GS}(z)\right|_{\lambda=0}=\frac{2}{a_{\|}\sqrt{\pi}}\exp{(-z^2/a^2_{\|})}
\sum_{n=0}^{N_{\rm f}/2-1}\frac{H^2_n(z/a_{\|})}{2^{n}n!}\,,
\end{equation}
given in terms of the Hermite polynomials $H_n(x)$ 
of degree $0\leq n\leq N_{\rm f}/2-1$. This 
density profile exhibits a shell structure characterized by $N_{\rm f}/2$ oscillations, whose 
origin lies in the fermionic statistical correlations: 
the occupation probability ${\cal P}(n)=\sum_\sigma \langle{\hat c}^{\dagger}_{n,\sigma}{\hat c}_{n,\sigma}\rangle$ of the $1D$ harmonic-oscillator states is unity for $0\leq n\leq N_{\rm f}/2-1$ and zero for $n\geq N_{\rm f}/2$. The existence of this sharp ``Fermi edge" 
is ultimately responsible for the bulk 
shell structure, which is analogous to the Friedel oscillations originating 
in a normal Fermi liquid from the sharply defined Fermi surface~\cite{Giuliani_and_Vignale}.
The occupation probabilities and the shell structure are expected 
to be strongly affected by many-body xc effects (see Sect.~\ref{sect:numerical_results} below).

(ii) 
In the limit $\lambda=+\infty$ the GS density profile should become that of a fully spin-polarized noninteracting 
Fermi gas,
\begin{equation}\label{eq:ferromagnetic}
\left.n_{\rm \scriptscriptstyle GS}(z)\right|_{\lambda=+\infty}=\frac{1}{a_{\|}\sqrt{\pi}}\exp{(-z^2/a^2_{\|})}
\sum_{n=0}^{N_{\rm f}-1}\frac{H^2_n(z/a_{\|})}{2^{n}n!}\,,
\end{equation}
exhibiting a shell structure characterized by $N_{\rm f}$ oscillations~\cite{noninteracting_1D}. This asymptotic property can be checked explicitly for $N_{\rm f}=2$ (see the Appendix) and originates from 
the fact that an infinitely strong repulsion between antiparallel-pseudospin atoms in $1D$ acts like the Pauli principle between parallel-pseudospin atoms~\cite{recati_2003,astrakharchik_2004}. The present formalism does not apply to such a strong coupling regime (note also that for $B>B^\star$ the value of $\lambda$ is bounded from above).

(iii) The main difference between the present BALDA scheme and 
the Thomas-Fermi approach is that in the latter~\cite{recati_2003,astrakharchik_2004,kecke_2005} the LDA is also 
used to approximate the noninteracting kinetic energy functional $T_{\rm s}[n_{\rm \scriptscriptstyle GS}(z)]$, which is written as
\begin{equation}
T_{\rm TF}[n_{\rm \scriptscriptstyle GS}(z)]=\int_{-\infty}^{+\infty}[n\kappa(n)]_{n\rightarrow n_{\rm \scriptscriptstyle GS}(z)}dz\,.
\end{equation}
The (Hohenberg-Kohn) Thomas-Fermi equation reads
\begin{equation}\label{eq:hk}
\left.\frac{\partial [n\kappa(n)]}{\partial n}\right|_{n\rightarrow n_{\rm \scriptscriptstyle GS}(z)}+V_{\rm KS}(z;[n_{\rm \scriptscriptstyle GS}(z)])={\rm constant}\,,
\end{equation}
the constant being fixed by normalization. The Thomas-Fermi profile misses the shell structure as well as atom tunnel beyond the Thomas-Fermi radius $Z_{\rm TF}$. In our approach, instead, $T_{\rm s}[n_{\rm \scriptscriptstyle GS}(z)]$ 
is treated exactly through the Kohn-Sham mapping
\begin{equation}
T_{\rm s}[n_{\rm \scriptscriptstyle GS}(z)]=-\frac{\hbar^2}{2m}\sum_{\alpha}\int_{-\infty}^{+\infty}\varphi^\star_\alpha(z)
\frac{\partial^2}{\partial z^2}\varphi_\alpha(z)dz\,,
\end{equation}	 
the single-particle orbitals $\varphi_\alpha(z)=\varphi_\alpha(z;[n_{\rm \scriptscriptstyle GS}(z)])$ being unique functionals of the GS density~\cite{dft,Giuliani_and_Vignale}.

\subsection{The exchange-correlation potential}

In what follows we propose two different ways to calculate the xc potential of the HGYM. 

\subsubsection{${\rm BALDA}/1$}
The potential $v^{\rm hom}_{\rm xc}(n,g_{\rm \scriptscriptstyle 1D})$ can be calculated by applying its definition~(\ref{eq:vxc_hom}) directly to Eqs.~(\ref{eq:energy_atom})-(\ref{eq:normalization}). 
It is easy to show that the xc potential of the HGYM is exactly given by the following equation,
\begin{equation}
v^{\rm hom}_{\rm xc}(n,g_{\rm \scriptscriptstyle 1D})=\frac{\varepsilon_{\rm b}}{2}
+2\nu\frac{\hbar^2 Q^2}{2m}\rho(Q)\frac{\partial Q}{\partial n}+2\pi\nu\int_{-Q}^{+Q}\frac{dk}{2\pi}\,\frac{\hbar^2 k^2}{2m}\frac{\partial \rho(k)}{\partial n}-\frac{\pi^2}{8}\hbar^2n^2-ng_{\rm \scriptscriptstyle 1D}\,,
\end{equation}
where $\partial_n Q$ and $\partial_n \rho$ satisfy the coupled BA equations
\begin{eqnarray}
\frac{\partial \rho(k)}{\partial n}&=&\frac{\nu}{2\pi\gamma n}\left[
{\cal K}\left(2(k-Q)/(\gamma n)\right)+{\cal K}\left(2(k+Q)/(\gamma n)\right)\right]
\frac{\partial Q}{\partial n}\nonumber\\
&+&\frac{\nu}{\gamma n}\int_{-Q}^{+Q}\frac{dk'}{2\pi}\,
{\cal K}\left(2(k-k')/(\gamma n)\right)\frac{\partial \rho(k')}{\partial n}
\end{eqnarray}
and
\begin{equation}
2\rho(Q)\frac{\partial Q}{\partial n}+\int_{-Q}^{+Q}\frac{\partial \rho(k)}{\partial n}dk=\frac{1}{\nu}\,.
\end{equation}
An accurate numerical solution of these coupled BA equations leads to the exact xc potential of the HGYM. 

The results for $n_{\rm \scriptscriptstyle GS}(z)$ that are obtained 
with $v^{\rm hom}_{\rm xc}(n,g_{\rm \scriptscriptstyle 1D})$ determined according to this route will be termed with the acronym ${\rm BALDA}/1$.

\subsubsection{${\rm BALDA}/2$}
As an alternative $v^{\rm hom}_{\rm xc}(n,g_{\rm \scriptscriptstyle 1D})$ can also be calculated from 
accurate analytical parametrizations of the GS energy of the HGYM, which incorporate exactly known limiting 
behaviors both at weak and strong coupling. This route will reduce the numerical effort and affords a test of the sensitivity of the results to the details of the implementation of the theory.

Let us introduce the Fermi wave number $k_F=\pi n/2$ and the Fermi energy $\varepsilon_F=\hbar^2 k^2_F/(2m)$. For repulsive interactions we find that the GS energy of the HGYM in units of the Fermi energy, 
$e=\varepsilon_{\rm \scriptscriptstyle GS}/\varepsilon_F$, can be very 
accurately parametrized by the simple formula
\begin{equation}\label{eq:fit_repulsive}
e(x>0)=\frac{4x^2/3+a_px+b_p}{x^2+c_px+d_p}\,,
\end{equation}
where $x=2\gamma/\pi$, $a_p=5.780126$, $b_p=-(8/9)\ln{2}+\pi a_p/4$, $c_p=(8/\pi)\ln{2}+3a_p/4$, and $d_p=3b_p$. Equation~(\ref{eq:fit_repulsive}) embodies the exact behaviors~\cite{astrakharchik_2004}
\begin{equation}\label{eq:exact_plus}
e(x)=\left\{
\begin{array}{ll}
1/3+x/\pi+...&{\rm for}\,x\rightarrow 0^+\\
4/3-32\ln{2}/(3\pi x)+...&{\rm for}\,x\rightarrow +\infty
\end{array}
\right.
\,.
\end{equation}
The fitting formula (\ref{eq:fit_repulsive}) is compared with the exact BA results in Fig.~\ref{fig:two} (top). 
Note that in the weak coupling limit our formula gives $e(x\rightarrow 0^+)=1/3+x/\pi-0.080x^2$, where the coefficient of the $x^2$ differs by about $4\%$ from the exact value determined by Magyar and Burke~\cite{burke_2004} 
by diagrammatic perturbation theory. On the other hand, the parametrization formula proposed by Magyar and Burke~\cite{burke_2004} incorporates exactly this second-order weak-coupling term, but violates the strong-coupling asymptotic result in Eq.~(\ref{eq:exact_plus}). In fact, the Magyar-Burke coefficient of the $1/x$ term, $c_{-1}=-1.829$, differs from the exact value $c_{-1}=-32\ln{2}/(3\pi)$ by about $22\%$.

Turning to the case of attractive interactions, we find the following parametrization formula 
to be very accurate:
\begin{equation}\label{eq:fit_attractive}
e(x<0)=\frac{1}{3}-\frac{|x|}{\pi}-{\cal A}(|x|)\frac{x^2}{4}
\end{equation}
where the function ${\cal A}(x)$, which modulates the amplitude of the strong-coupling term $-x^2/4$, 
is given by
\begin{equation}
{\cal A}(x)=\frac{x^2+a_m x+b_m}{x^2+c_m x+ d_m}
\end{equation}
with $a_m=-0.331117$, $b_m=0.458183$, $c_m=a_m+4/\pi$, and $d_m=4a_m/\pi+b_m+16/\pi^2-1$.
Equation~(\ref{eq:fit_attractive}) embodies the exact asymptotic behaviors~\cite{astrakharchik_2004}
\begin{equation}
e(x)=\left\{
\begin{array}{ll}
1/3+x/\pi+...&{\rm for}\,x\rightarrow 0^-\\
-x^2/4+1/12+...&{\rm for}\,x\rightarrow -\infty
\end{array}
\right.
\,.
\end{equation}
Equation~(\ref{eq:fit_attractive}) is compared with the exact BA results in Fig.~\ref{fig:two} (bottom). 

The xc potential can be calculated analytically using its definition in Eq.~(\ref{eq:vxc_hom}) applied to Eqs.~(\ref{eq:fit_repulsive}) or~(\ref{eq:fit_attractive}). The results for $n_{\rm \scriptscriptstyle GS}(z)$ that are obtained with $v^{\rm hom}_{\rm xc}(n,g_{\rm \scriptscriptstyle 1D})$ determined according to this parametrization procedure 
will be termed with the acronym ${\rm BALDA}/2$.

\section{Numerical Results}
\label{sect:numerical_results}

We proceed to illustrate our main numerical results, which are summarized 
in Figs.~\ref{fig:three}-\ref{fig:eight}. In Fig.~\ref{fig:three} we show the GS density profile of a Fermi gas with $N_{\rm f}=10$ atoms at $\lambda=+2$ and $-2$. Repulsive interactions depress the amplitude of the shell structure, while attractive interactions enhance the oscillations of the profile leading to an ADW with $N_{\rm f}/2$ distinct maxima related to the formation of $N_{\rm f}/2$ spin pairs. The two BALDA schemes that we have proposed are in excellent agreement with each other, showing no visible difference on the scale of the figure. In Fig.~\ref{fig:four} we report a summary of our ${\rm BALDA}/1$ results for the GS density profiles of a paramagnetic Fermi gas with $N_{\rm f}=10$ atoms for increasing repulsive or attractive interactions.

The shell structure is also sensitive to the system size, with larger clouds tending to have a relatively weaker structure. In Fig.~\ref{fig:five} we show the GS density profile of a cloud with $N_{\rm f}=20$ atoms at $\lambda=+2$ and $-2$. Comparing this figure with Fig.~\ref{fig:three} it is clear 
that the amplitude of the oscillations is decreasing with increasing $N_{\rm f}$. 
In particular for $N_{\rm f}=50$ atoms at $\lambda=+2$ (see Fig.~\ref{fig:six}, top) 
the Thomas-Fermi results are a good representation of the actual density profile, 
except at the edges of the cloud. On the other hand, for the same system size 
attractive interactions at $\lambda=-2$ lead to a still clearly visible ADW, though this is absent in the Thomas-Fermi theory (see Fig.~\ref{fig:six}, bottom).

Finally, the problem of two atoms with opposite pseudospins in Q$1D$ harmonic confinement is exactly solvable (see Appendix). {\it A priori} we do not expect an LDA approach to be applicable to such a small system, but from Fig.~\ref{fig:seven} it is seen that the BALDA scheme still yields some reasonable results for both repulsive and attractive interactions. For strong repulsive interactions it is not able to reproduce the formation of a hole at the center of the trap (see the discussion under point (ii) of the previous Section), while 
in the case of strong attractive interactions it overestimates the value of the density in the same region (see Fig.~\ref{fig:seven}). The method is nevertheless usefully reliable for the GS energy over a wide range of values of $\lambda$ (see Fig.~\ref{fig:eight}): for instance, at $\lambda=-30$ we find ${\cal E}^{\rm BALDA}_{\rm \scriptscriptstyle GS}/(N_{\rm f}\hbar\omega_{\|})=-224.810$ as compared to the exact value ${\cal E}_{\rm \scriptscriptstyle GS}/(N_{\rm f}\hbar\omega_{\|})=-224.499$, and at $\lambda=+30$ we find ${\cal E}^{\rm BALDA}_{\rm \scriptscriptstyle GS}/(N_{\rm f}\hbar\omega_{\|})=+1.01$ as compared to the exact value ${\cal E}_{\rm \scriptscriptstyle GS}/(N_{\rm f}\hbar\omega_{\|})=+0.975$.

\section{Summary and Conclusions}
\label{sect:conclusions}

In summary, we have presented a novel Kohn-Sham DFT study of two-component Fermi gases 
with repulsive or attractive intercomponent interactions 
in Q$1D$ harmonic traps. The present BALDA theory, which is expected to be accurate 
in a weak-to-intermediate range of coupling strength, provides a quantitative microscopic understanding 
of how many-body exchange and correlations modify the bulk shell structure of the ground-state density profile. 
Repulsive intercomponent interactions depress the amplitude of the shell structure while 
attractive interactions stabilize atomic-density waves through Luther-Emery spin pairing. Such atomic-density waves should be observable in gaseous clouds containing of the order of up to $100$ atoms, a suitable experimental technique being the search for satellites in the elastic diffraction pattern as discussed in Ref.~\cite{gao_2005}. It would also be important to re-examine numerically these GS exchange and correlation properties in relatively small systems with $N_{\rm f}\lesssim 10$ atoms by exact-diagonalization or Quantum Monte Carlo methods.

The present work can be extended in several directions. For instance, it would be interesting to generalize the present scheme to the composite-boson region $B<B^\star$ in order to have a DFT treatment of the BCS-BEC crossover in the presence of axial confinement, generalizing the theories by Fuchs {\it et al.}~\cite{fuchs_2004} and by Tokatly~\cite{tokatly_2004}. Secondly, it would be interesting to study dynamical phenomena such as spin-charge separation in these strongly correlated gases using time-dependent DFT and/or current-DFT~\cite{vk_1996,Giuliani_and_Vignale}, instead of resorting to the inhomogeneous Tomonaga-Luttinger liquid model~\cite{recati_2003,kecke_2005}. From a more formal DFT viewpoint, a functional better than in Eq.~(\ref{eq:balda}) is desirable and necessary to deal with the strong coupling regime.

\acknowledgments
This work was partially supported by MIUR through the PRIN2003 Program. G. X. and M. P. T. 
acknowledge the hospitality of the 
Condensed Matter Section of the Abdus Salam International Center for Theoretical Physics, where
part of the calculations was done. It is a pleasure to thank Dr. P. Capuzzi, 
Prof. Jianhui Dai, and Dr. B. Davoudi for many useful discussions. 

\appendix*
\section{Exact solution of the two-atom problem}

The problem of two antiparallel-spin fermions interacting with a zero-range delta-function potential in $1D$ 
is exactly solvable thanks to the separation of centre-of-mass and relative variables, which is 
allowed by the harmonic trapping potential. Performing a canonical transformation to centre-of-mass $(Z=(z_1+z_2)/2,P=p_1+p_2)$ and relative $(z_{\rm rel}=z_1-z_2,p=(p_1-p_2)/2)$ coordinates and momenta, 
the Hamiltonian can be written as
${\cal H}={\cal H}_{\rm CM}(Z,P)+{\cal H}_{\rm rel}(z_{\rm rel},p)$. Here, 
the centre-of-mass Hamiltonian ${\cal H}_{\rm CM}=P^2/(2M)+M\omega^2_{\|}Z^2/2$ describes 
a free particle of mass $M=2m$ in a $1D$ harmonic oscillator, while the relative-motion Hamiltonian ${\cal H}_{\rm rel}=p^2/(2\mu)+{\cal V}(z_{\rm rel})$ 
describes a free particle of mass $\mu=m/2$ in the potential ${\cal V}(z_{\rm rel})=\mu\omega^2_{\|}z^2_{\rm rel}/2+g_{\rm 1D}\delta(z_{\rm rel})$. 

The spatial part of the GS wavefunction can thus be written as
\begin{equation}
\Psi_{\rm GS}(z_1,z_2)={\cal N}\exp{(-Z^2/a_{\|}^2)}\varphi^{(0)}_{\rm rel}(z_{\rm rel})\,,
\end{equation}
where ${\cal N}$ is a normalization constant and $\varphi^{(0)}_{\rm rel}(z_{\rm rel})$ is the GS wavefunction of the relative-motion problem with energy $\varepsilon_0$. 
Introducing the dimensionless coordinate ${\bar z}_{\rm rel}=z_{\rm rel}/a_{\|}$, 
the Schr\"odinger equation for the relative motion reads
\begin{equation}\label{eq:relative}
\left[-\frac{d^2}{d {\bar z}^2_{\rm rel}}+\frac{1}{4}{\bar z}^2_{\rm rel}+
\lambda\delta({\bar z}_{\rm rel})\right]\varphi^{(n)}_{\rm rel}({\bar z}_{\rm rel})={\bar \varepsilon}_n
\varphi^{(n)}_{\rm rel}({\bar z}_{\rm rel})\,,
\end{equation}
where ${\bar \varepsilon}_n=\varepsilon_n/(\hbar\omega_{\|})$. 
Due to the antisymmetric (spin-singlet) nature of the spinorial part of the GS wavefunction, 
we need to search for the lowest ($n=0$) even eigensolution of Eq.~(\ref{eq:relative}). 

The singular delta-function term in Eq.~(\ref{eq:relative}) imposes a cusp on the wavefunctions at the origin,
\begin{equation}\label{eq:cusp}
\lim_{{\bar z}_{\rm rel}\rightarrow 0^+}\frac{\partial\varphi^{(n)}_{\rm rel}({\bar z}_{\rm rel})}{\partial {\bar z}_{\rm rel}}-\lim_{{\bar z}_{\rm rel}\rightarrow 0^-}
\frac{\partial\varphi^{(n)}_{\rm rel}({\bar z}_{\rm rel})}{\partial {\bar z}_{\rm rel}}=
\lambda\varphi^{(n)}_{\rm rel}({\bar z}_{\rm rel}=0)\,.
\end{equation}
Eq.~(\ref{eq:relative}) is then recognized to be the differential equation that defines the Parabolic Cylinder Functions~\cite{abram}. The even solutions with the proper asymptotic behavior are~\cite{busch_huyel_2003,abram} 
\begin{equation}
\varphi^{(n)}_{\rm rel}({\bar z}_{\rm rel})=
D_{{\bar \varepsilon}_n-1/2}(|{\bar z}_{\rm rel}|)\,,
\end{equation}
$D_a(x)$ being a Whittaker function. Using the following properties of the Whittaker function~\cite{abram},
\begin{equation}
\varphi^{(n)}_{\rm rel}({\bar z}_{\rm rel}=0)
=\frac{\sqrt{\pi}}{2^{-{\bar \varepsilon}_n/2+1/4}\Gamma(3/4-{\bar \varepsilon}_n/2)}
\end{equation}
and
\begin{equation}
\lim_{{\bar z}_{\rm rel}\rightarrow 0^+}\frac{\partial\varphi^{(n)}_{\rm rel}({\bar z}_{\rm rel})}{\partial {\bar z}_{\rm rel}}=-\frac{\sqrt{\pi}}{2^{-{\bar \varepsilon}_n/2-1/4}\Gamma(1/4-{\bar \varepsilon}_n/2)}
\end{equation}
together with
\begin{equation}
\lim_{{\bar z}_{\rm rel}\rightarrow 0^-}\frac{\partial\varphi^{(n)}_{\rm rel}({\bar z}_{\rm rel})}{\partial {\bar z}_{\rm rel}}=-\lim_{{\bar z}_{\rm rel}\rightarrow 0^+}\frac{\partial\varphi^{(n)}_{\rm rel}({\bar z}_{\rm rel})}{\partial {\bar z}_{\rm rel}}\,,
\end{equation}
it is easy to show that Eq.~(\ref{eq:cusp}) becomes the trascendental equation
\begin{equation}\label{eq:energy_exact} 
\frac{\Gamma(3/4-{\bar \varepsilon}_n/2)}{\Gamma(1/4-{\bar \varepsilon}_n/2)}=-\frac{\lambda}{2\sqrt{2}}\,.
\end{equation}
Here $\Gamma(x)$ is the Euler Gamma function. This equation implicitly defines the function ${\bar \varepsilon}_n(\lambda)$. The l.h.s. of Eq.~(\ref{eq:energy_exact}) is shown in Fig.~\ref{fig:nine} as a function of $-{\bar \varepsilon}_n$. The GS energy per atom is given by ${\cal E}_{\rm \scriptscriptstyle GS}/(N_{\rm f}\hbar\omega_{\|})=1/4+{\bar \varepsilon}_0(\lambda)/2$ and is shown in Fig.~\ref{fig:eight} as a solid line. 

Some limiting behaviors of the function ${\bar \varepsilon}_0(\lambda)$ can be established analytically from the properties of the Gamma function. We find
\begin{equation}
{\bar \varepsilon}_0(\lambda \rightarrow 0)=\frac{1}{2}+\frac{\lambda}{\sqrt{2\pi}}-\frac{\lambda^2}{2\pi}\ln{2}+...
\end{equation}
in the weak coupling limit,
\begin{equation}
{\bar \varepsilon}_0(\lambda \rightarrow +\infty)=\frac{3}{2}-2\sqrt{\frac{2}{\pi}}\frac{1}{\lambda}-\frac{8}{\pi}(\ln{2}-1)\frac{1}{\lambda^2}+...
\end{equation}
in the strong repulsion limit, and
\begin{equation}
{\bar \varepsilon}_0(\lambda\rightarrow -\infty)=-\frac{\lambda^2}{4}-\frac{1}{2\lambda^2}+\frac{9}{2\lambda^6}+...
\end{equation}
in the strong attraction limit.

The GS density profile can be found from 
\begin{equation}
n_{\rm GS}(z)=\int_{-\infty}^{+\infty}dz'\,|\Psi_{\rm GS}(z,z')|^2\,.
\end{equation}
The normalization constant ${\cal N}$ is chosen according to $\int_{-\infty}^{+\infty}dz\,n_{\rm GS}(z)=2$, {\it i.e.}
\begin{equation}
{\cal N}^2=\frac{2^{3/2}/(\sqrt{\pi}a_{\|})}{\displaystyle \int_{-\infty}^{+\infty} dz_{\rm rel} |\varphi^{(0)}_{\rm rel}(z_{\rm rel})|^2}
\end{equation}
where~\cite{grad}
\begin{eqnarray}\label{eq:relative_normalization}
\int_{-\infty}^{+\infty}dz_{\rm rel} |\varphi^{(n)}_{\rm rel}(z_{\rm rel})|^2=\sqrt{\frac{\pi}{2}}a_{\|}\,\frac{\psi(3/4-{\bar \varepsilon}_n/2)-\psi(1/4-{\bar \varepsilon}_n/2)}{\Gamma(1/2-{\bar \varepsilon}_n)}\,,
\end{eqnarray}
with $\psi(x)=d\ln{\Gamma(x)}/dx$. The relation (\ref{eq:relative_normalization}) holds unless ${\bar \varepsilon}_n=n+1/2$ with $n=0,1,2,...$, when one has to use the result~\cite{grad}
\begin{equation}
\int_0^{+\infty}d{\bar z}_{\rm rel} |D_n({\bar z}_{\rm rel})|^2=\sqrt{\frac{\pi}{2}}\,n!\,.
\end{equation}
For instance, in the case $\lambda=+\infty$ we find ${\bar \varepsilon}_0=3/2$, $\varphi^{(0)}_{\rm rel}({\bar z}_{\rm rel})=|{\bar z}_{\rm rel}|\exp{(-{\bar z}^2_{\rm rel}/4)}$, ${\cal N}^2=2/(\pi a^2_{\|})$ and 
\begin{eqnarray}
\left.n_{\rm GS}(z)\right|_{\lambda=+\infty}=\left[1+2(z/a_{\|})^2\right]\,\frac{1}{\sqrt{\pi}a_{\|}}\exp{(-z^2/a^2_{\|})}\,.
\end{eqnarray}
This result has form given in Eq.~(\ref{eq:ferromagnetic}).

\begin{figure}
\begin{center}
\includegraphics[width=1.00\linewidth]{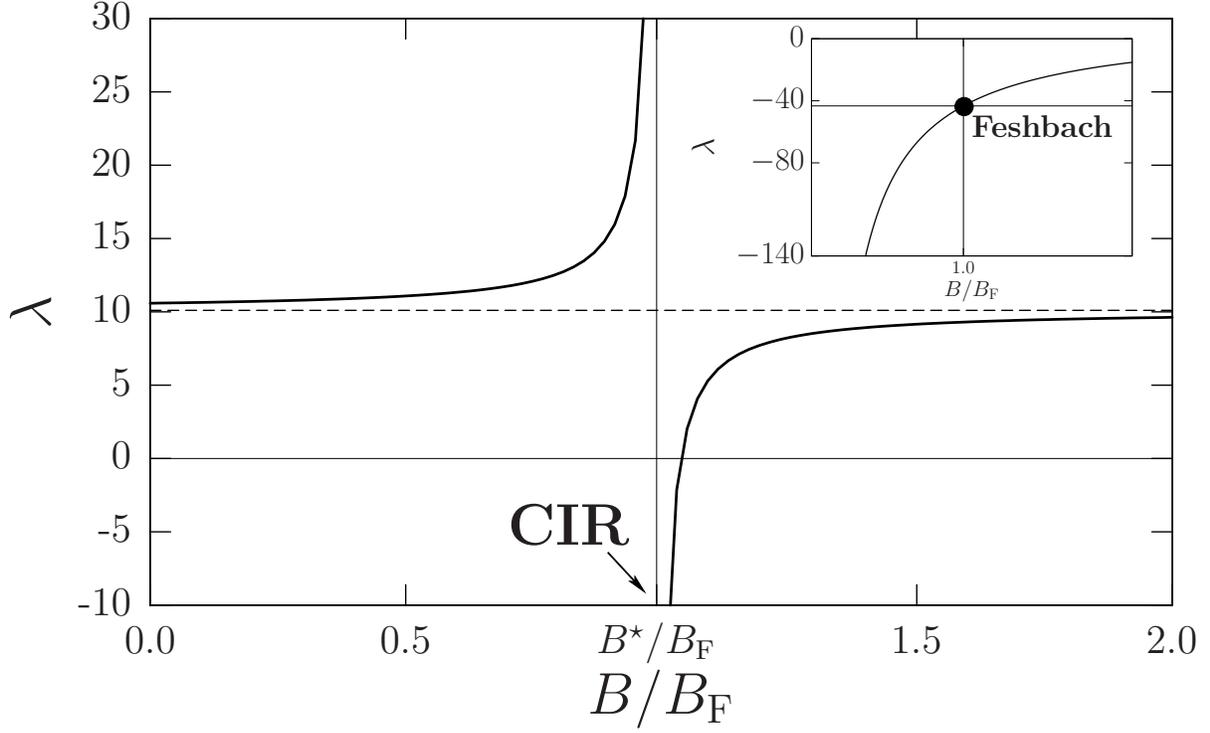}
\caption{The dimensionless coupling parameter $\lambda$ as a function of the magnetic field $B$ (in units of the Feshbach resonance field $B_{\rm F}$). Here we have chosen the following values for the relevant parameters: $m=6.642\times10^{-26}\,{\rm Kg}$ (mass of a $^{40}{\rm K}$ atom), $\omega_{\perp}=2\pi\,\times 100\,{\rm kHz}$ and $\omega_{\|}=2\pi\,\times 200\,{\rm Hz}$ (the anisotropy parameter of the trap is $2\times10^{-3}$), $B_{\rm F}=202.1\,{\rm G}$, $\delta B=7.8\,{\rm G}$, and $a_{\rm bg}=174$ Bohr radii. For these parameters $B^\star=0.991B_{\rm F}$. At the $3D$ Feshbach resonance $\lambda^{\rm F}=-43.309$ (see inset).\label{fig:one}}
\end{center}
\end{figure}

\begin{figure}
\begin{center}
\tabcolsep=0 cm
\begin{tabular}{c}
\includegraphics[width=1.00\linewidth]{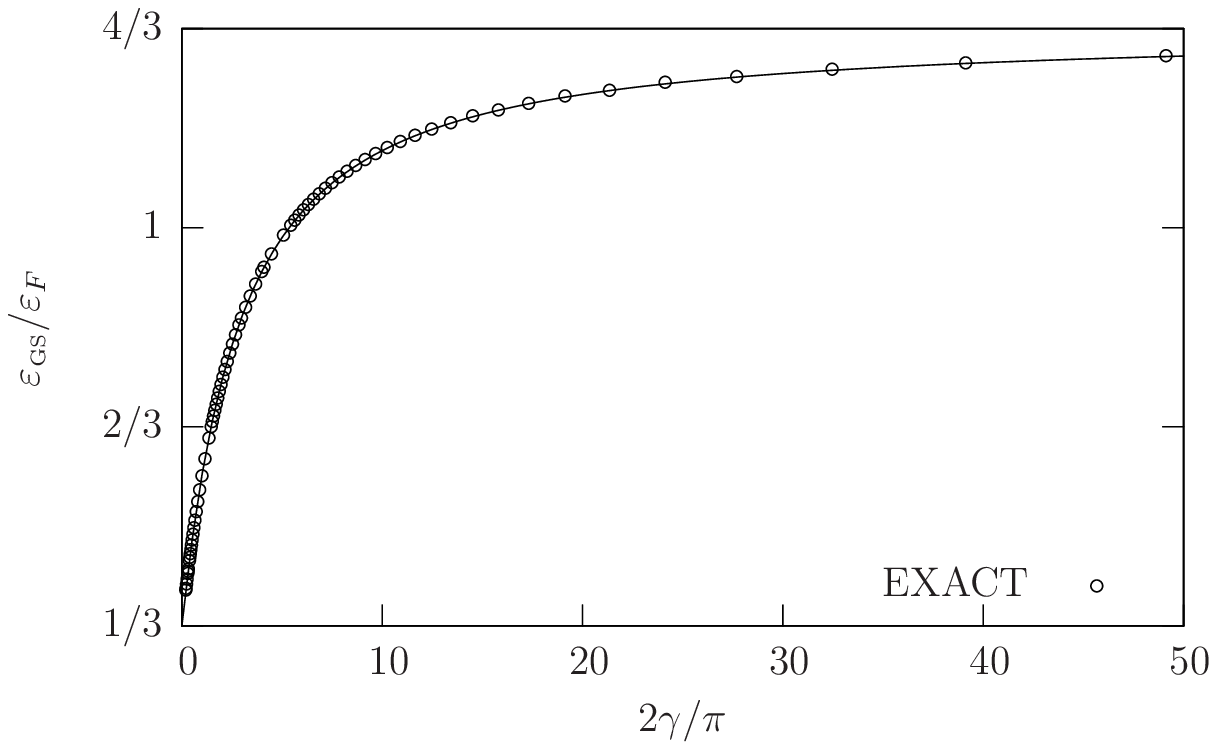}\\
\includegraphics[width=1.00\linewidth]{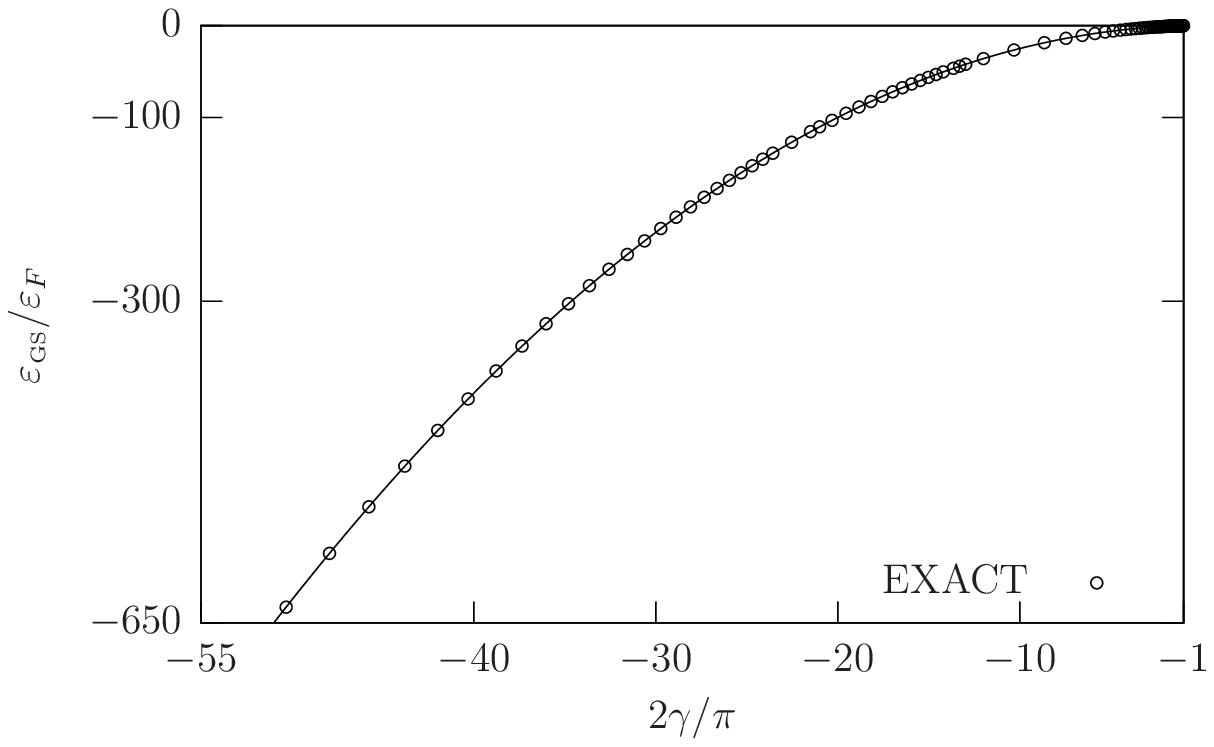}
\end{tabular}
\caption{Ground-state energy $\varepsilon_{\rm \scriptscriptstyle GS}(n,g_{\rm \scriptscriptstyle 1D})$ 
of the HGYM (per particle and in units of the Fermi energy $\varepsilon_F$) as a function of the coupling strength $2\gamma/\pi$ for a paramagnetic Fermi gas with repulsive interactions (top) and attractive interactions (bottom). The exact results, obtained from the solution of the BA equations (\ref{eq:energy_atom})-(\ref{eq:normalization}), are compared with the fitting formulae in Eq.~(\ref{eq:fit_repulsive}) and in Eq.~(\ref{eq:fit_attractive}) (solid lines).\label{fig:two}}
\end{center}
\end{figure}

\begin{figure}
\begin{center}
\includegraphics[width=1.00\linewidth]{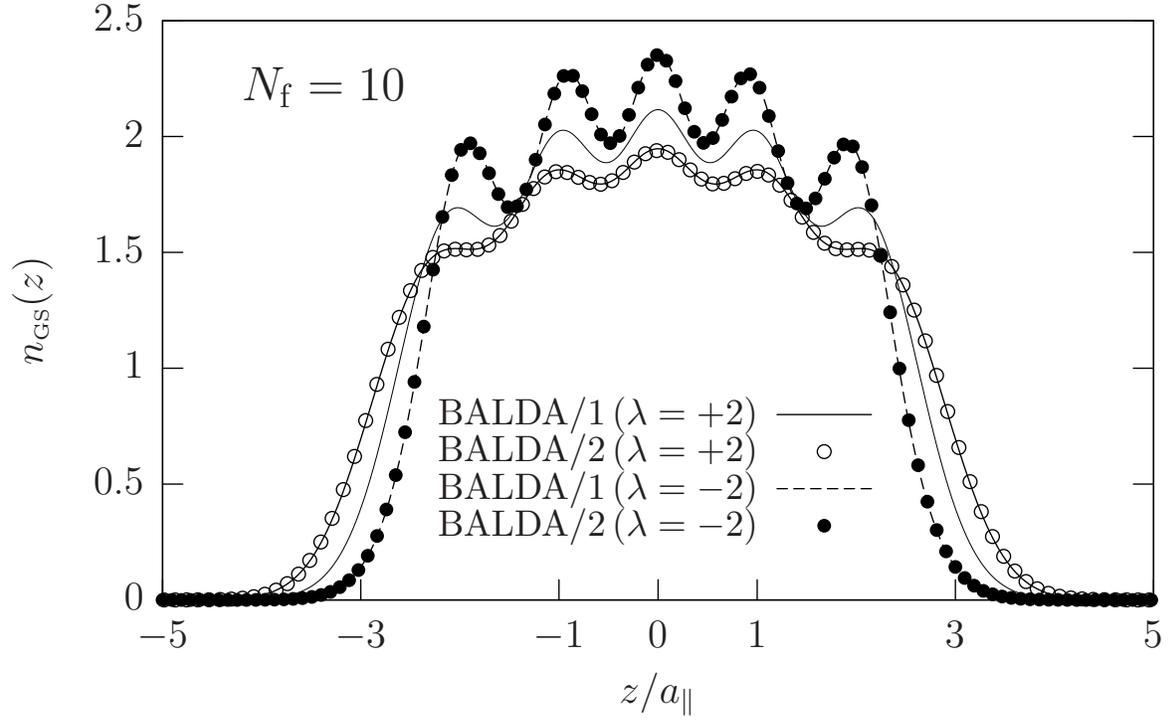}
\end{center}
\caption{Density profile $n_{\rm GS}(z)$ (in units of $a_{\|}^{-1}$) as a function of $z/a_{\|}$ 
for a paramagnetic Fermi gas with $N_{\rm f}=10$ atoms at $\lambda=+2$ and $-2$. 
The results of the ${\rm BALDA}/1$ scheme are compared with those of the ${\rm BALDA}/2$ scheme.
The thin solid line corresponds to the noninteracting $\lambda=0$ case.\label{fig:three}}
\end{figure}

\begin{figure}
\begin{center}
\tabcolsep=0 cm
\begin{tabular}{c}
\includegraphics[width=1.00\linewidth]{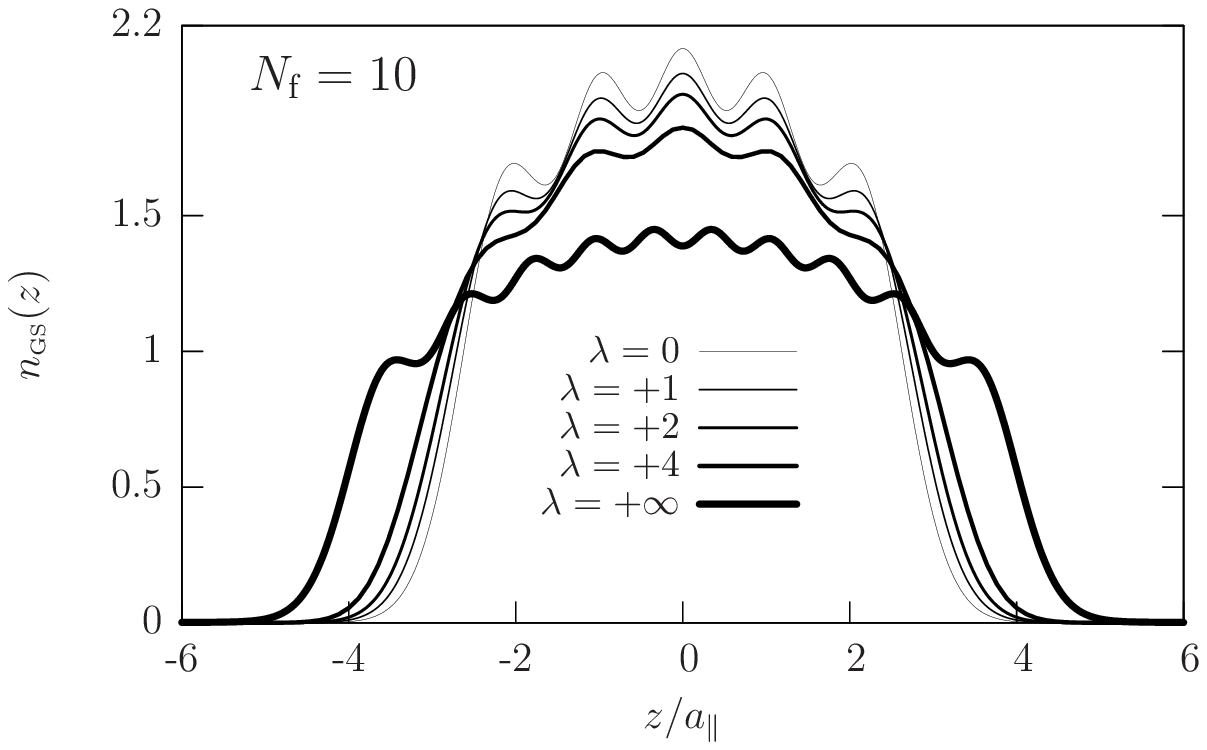}\\
\includegraphics[width=1.00\linewidth]{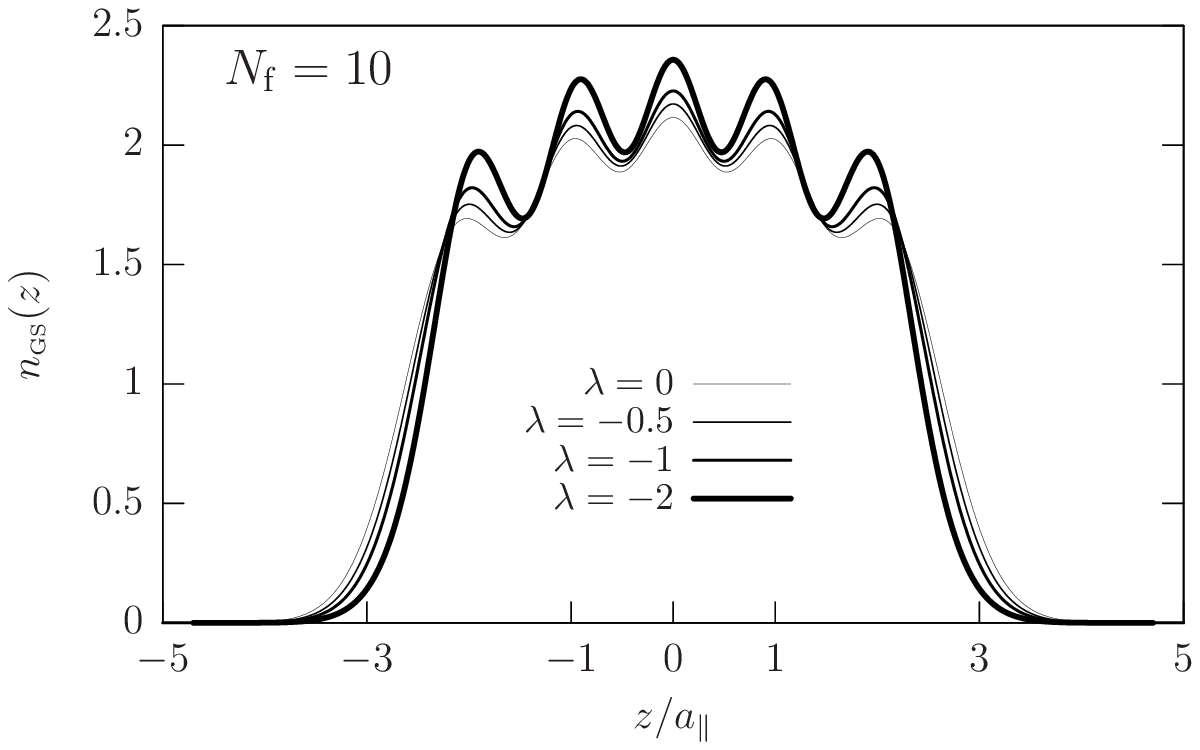}
\end{tabular}
\caption{Evolution of the density profile $n_{\rm GS}(z)$ (in units of $a_{\|}^{-1}$) with increasing $\lambda$ in the ${\rm BALDA}/1$ scheme, for a paramagnetic Fermi gas of $N_{\rm f}=10$ atoms with repulsive interactions (top) and attractive interactions (bottom). The curve at $\lambda=+\infty$ is the theoretical result given in Eq.~(\ref{eq:ferromagnetic}).\label{fig:four}}
\end{center}
\end{figure}

\begin{figure}
\begin{center}
\includegraphics[width=1.00\linewidth]{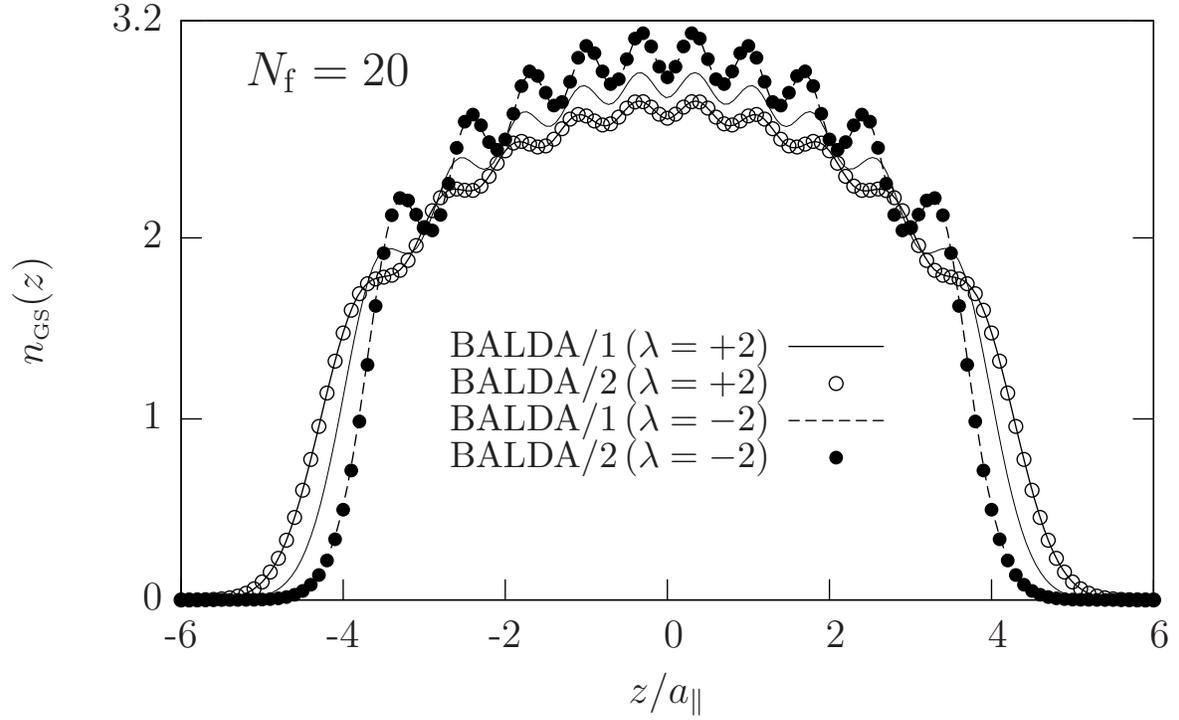}
\caption{Density profile $n_{\rm GS}(z)$ (in units of $a_{\|}^{-1}$) as a function of $z/a_{\|}$ 
for a paramagnetic Fermi gas with $N_{\rm f}=20$ atoms at $\lambda=+2$ and $-2$. 
The thin solid line corresponds to the noninteracting $\lambda=0$ case.\label{fig:five}}
\end{center}
\end{figure}

\begin{figure}
\begin{center}
\tabcolsep=0 cm
\begin{tabular}{c}
\includegraphics[width=1.00\linewidth]{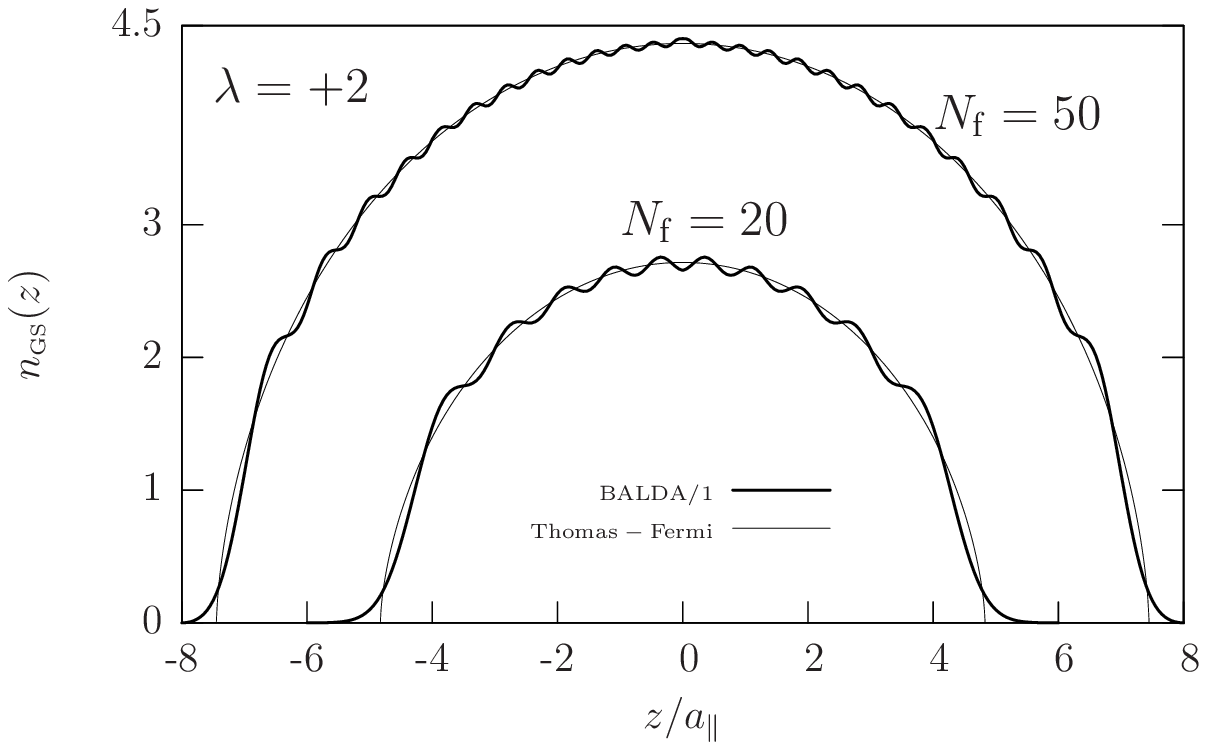}\\
\includegraphics[width=1.00\linewidth]{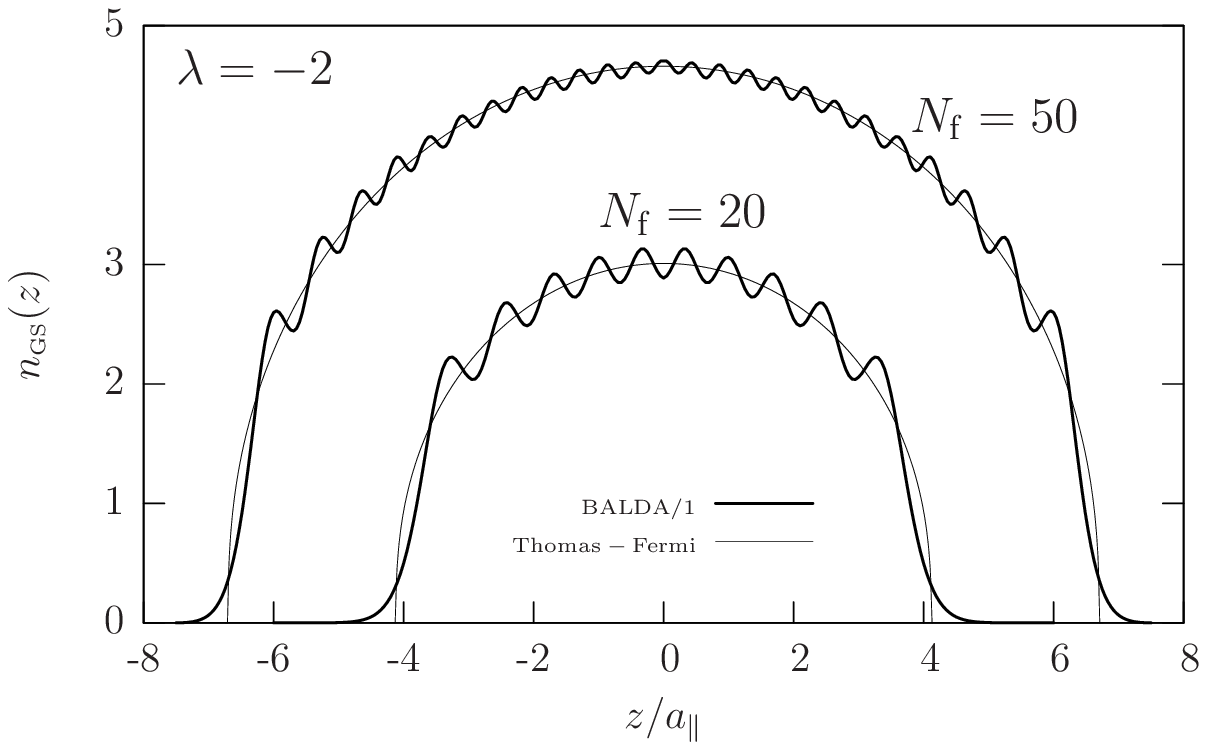}
\end{tabular}
\caption{Density profile $n_{\rm GS}(z)$ (in units of $a_{\|}^{-1}$) as a function of $z/a_{\|}$ 
for paramagnetic Fermi gases with $N_{\rm f}=20$ and $50$ atoms at $\lambda=+2$ (top) and $\lambda=-2$ (bottom). 
The results of the ${\rm BALDA}/1$ scheme are compared with the Thomas-Fermi results.\label{fig:six}}
\end{center}
\end{figure}

\begin{figure}
\begin{center}
\tabcolsep=0 cm
\begin{tabular}{c}
\includegraphics[width=1.00\linewidth]{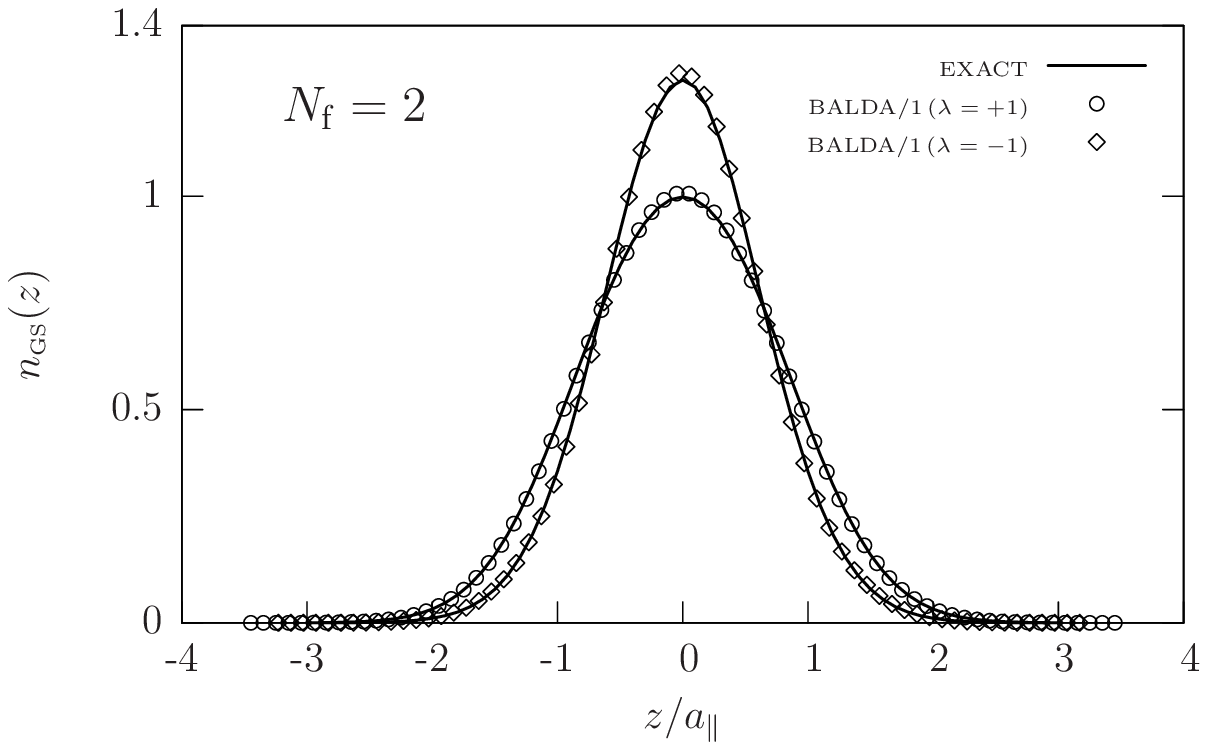}\\
\includegraphics[width=1.00\linewidth]{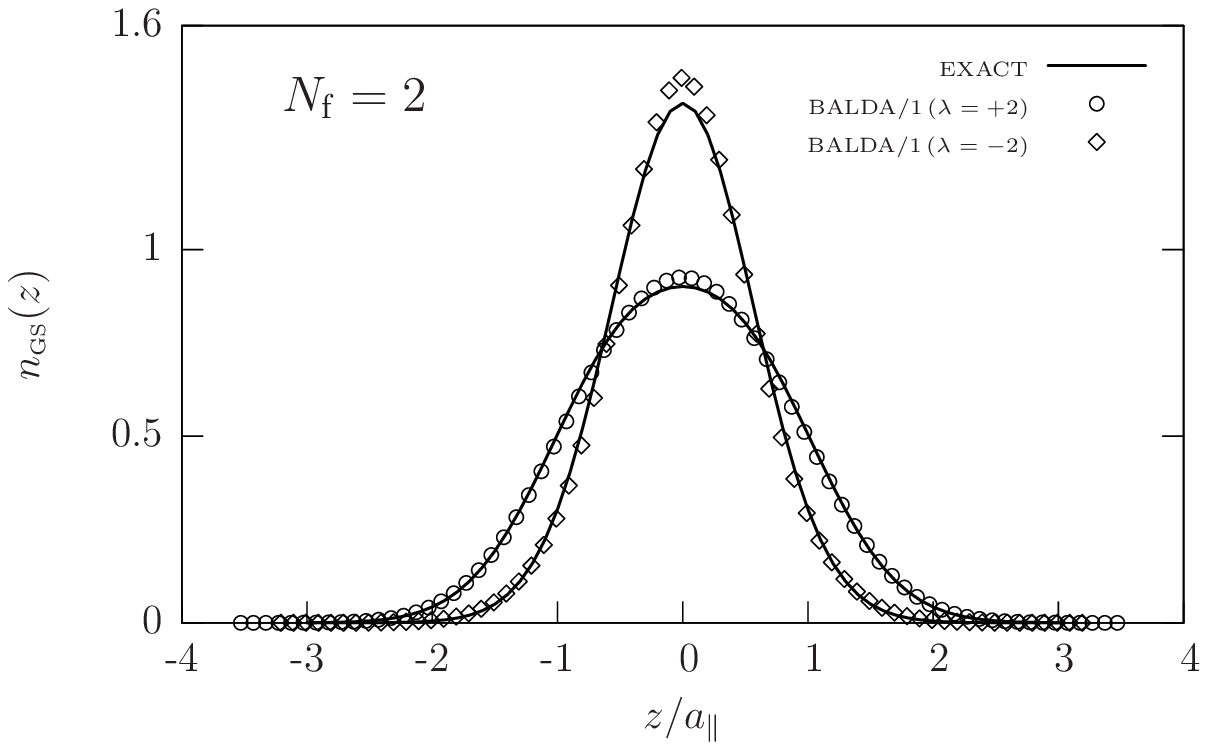}\\
\includegraphics[width=1.00\linewidth]{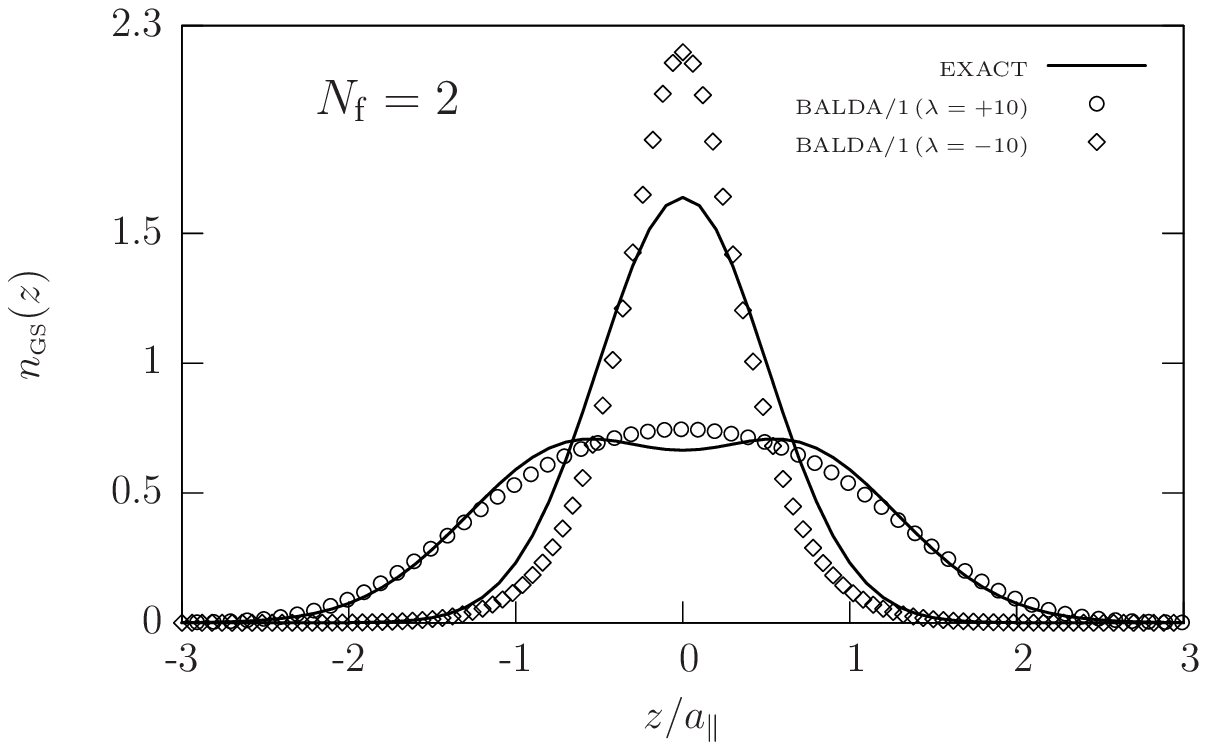}
\end{tabular}
\caption{Density profile $n_{\rm GS}(z)$ (in units of $a_{\|}^{-1}$) as a function of $z/a_{\|}$ 
for two Fermi atoms with opposite pseudospins at $\lambda=+1$ and $-1$ (top), $\lambda=+2$ and $-2$ (middle), and $\lambda=+10$ and $-10$ (bottom). The results of the ${\rm BALDA}/1$ scheme are compared with the exact results.\label{fig:seven}}
\end{center}
\end{figure}

\begin{figure}
\begin{center}
\includegraphics[width=1.00\linewidth]{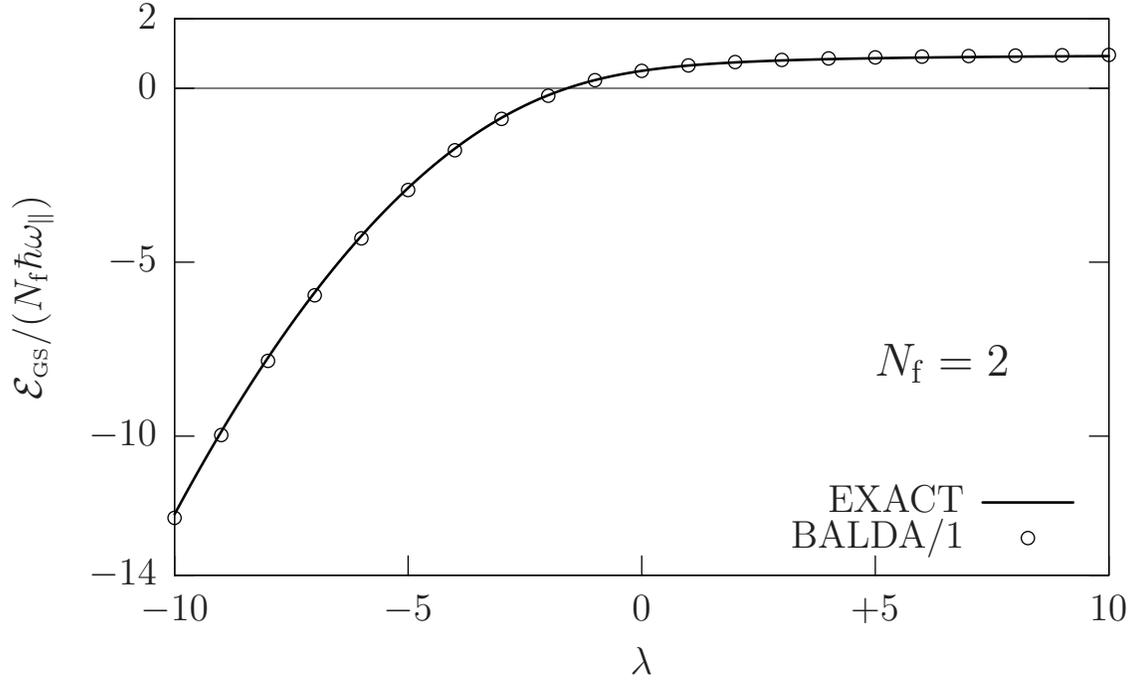}
\end{center}
\caption{Ground-state energy per atom ${\cal E}_{\rm \scriptscriptstyle GS}/N_{\rm f}$ 
(in units of $\hbar\omega_{\|}$) as a function of $\lambda$ 
for two Fermi atoms with opposite pseudospins. The results of the ${\rm BALDA}/1$ scheme 
are compared with the exact results.\label{fig:eight}}
\end{figure}

\begin{figure}
\begin{center}
\includegraphics[width=1.00\linewidth]{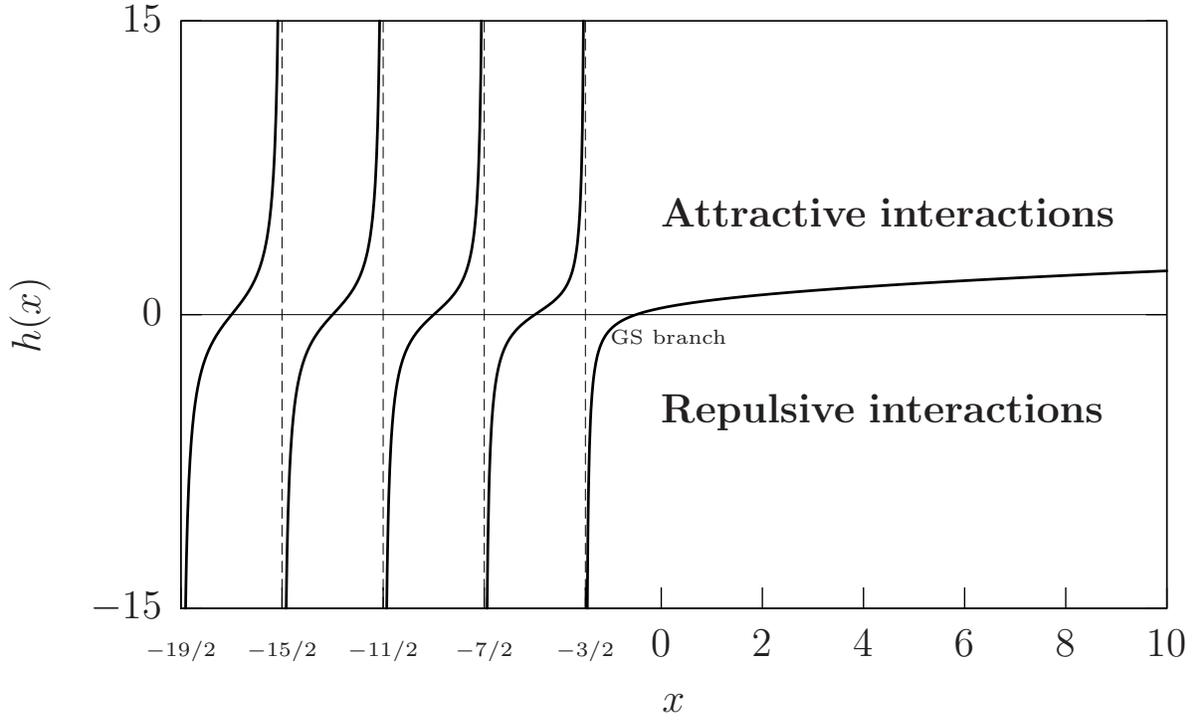}
\end{center}
\caption{A plot of the function $h(x)=\Gamma(3/4+x/2)/\Gamma(1/4+x/2)$. $h(x)$ has zeroes 
at $x=-2n-1/2$ and poles at $x=-2n-3/2$, with $n=0,1,2,...$. In order to find the GS energy of the $2$-atoms system one has to find the intersections of horizontal lines with the branch of $h(x)$ in the range $-3/2\leq x<+\infty$. The upper (lower) half-plane is relevant for attractive (repulsive) interactions.\label{fig:nine}}
\end{figure}

\end{document}